\documentclass[aps,twocolumn,amsmath,amssymb,showpacs,prb]{revtex4}
\usepackage{epsf}
\usepackage{graphicx}

\begin{document}

\title{Identifying the Background Signal in ARPES of High Temperature 
Superconductors}
\author{
        A. Kaminski,$^{1,2}$
        S. Rosenkranz,$^{2,3}$
        H. M. Fretwell,$^{1}$
        J. Mesot,$^{4}$
        M. Randeria,$^{5}$
        J. C. Campuzano,$^{2,3}$
        M. R. Norman,$^{3}$
        Z. Z. Li,$^{6}$
        H. Raffy,$^{6}$
        T. Sato,$^7$
        T. Takahashi $^7$ and
        K. Kadowaki$^8$
       }
\affiliation{
         (1) Department of Physics,University of Wales Swansea, Swansea SA2 8PP, UK\\
         (2) Department of Physics, University of Illinois at Chicago, Chicago, IL 60607\\
         (3) Materials Sciences Division, Argonne National Laboratory, Argonne, IL 60439 \\
         (4) Laboratory for Neutron Scattering, ETH Zurich and PSI Villigen, CH-5232 Villigen PSI, Switzerland\\
         (5) Tata Institute of Fundamental Research, Homi Bhabha Road, Mumbai 400005, India\\
         (6) Laboratoire de Physique des Solides, Universit\'{e} Paris-Sud, 91405 Orsay, France\\
         (7) Department of Physics, Tohoku University, Sendai 980-8578, Japan\\
         (8) Institute of Materials Science, University of Tsukuba, Ibaraki 305-3573, Japan\\
         }
\date{\today}
\begin{abstract}
One of the interesting features of the photoemission spectra of the high temperature cuprate superconductors is the presence of a large signal (referred to as the ``background'') in the unoccupied region of the Brillouin zone. Here we present data indicating that the origin of this signal is extrinsic and is most likely due to strong scattering of the photoelectrons. We also present an analytical method that can be used to subtract the background signal.
\end{abstract}

\maketitle

Angle resolved photoemission spectroscopy (ARPES) has provided a unique insight into the electronic structure of the high temperature superconductors \cite{RMP}. Information from ARPES about the Fermi surface, the symmetry of the superconducting gap and pseudogap, as well as the many body interactions and their energy scales, have stimulated a number of theoretical papers. As a result of recent technological advances in electron optics, even more precise information is now being obtained. In order to proceed with a detailed analysis of such data, however, one needs to understand some subtle properties of the ARPES spectra and be able to separate intrinsic features from extrinsic ones. 

Early ARPES experiments on high temperature superconductors revealed an unusually large signal outside of the occupied region of the Brillouin zone - sometimes referred to as the ``background". Its origin has been debated for a long time, with some suggesting it to be an intrinsic property (incoherent part of the spectral function), while others regard it as an extrinsic effect \cite{BACK}.  If the background signal is extrinsic (e.g.~due to photoelectron scattering), it would be a source of contamination of the signal inside of the Fermi surface as well. Answering this question is thus quite important, since it impacts our understanding of the nature of the many body interactions in the cuprates as well as a quantitative analysis of ARPES data.

Here we present ARPES data collected at the Synchrotron Radiation Center undulator 4m NIM beamline using both Scienta SES200 and SES50 analyzers. The optimally doped thin film samples of Bi2212 were grown using an RF sputtering technique while the optimally doped single crystal samples were grown in a floating zone furnace. To obtain quantitative information about the intensity, we divide each spectra by the acquisition time and the photon flux - measured by a Ni mesh at the entrance of the experimental chamber.

\begin{figure}
\includegraphics[width=3.5in]{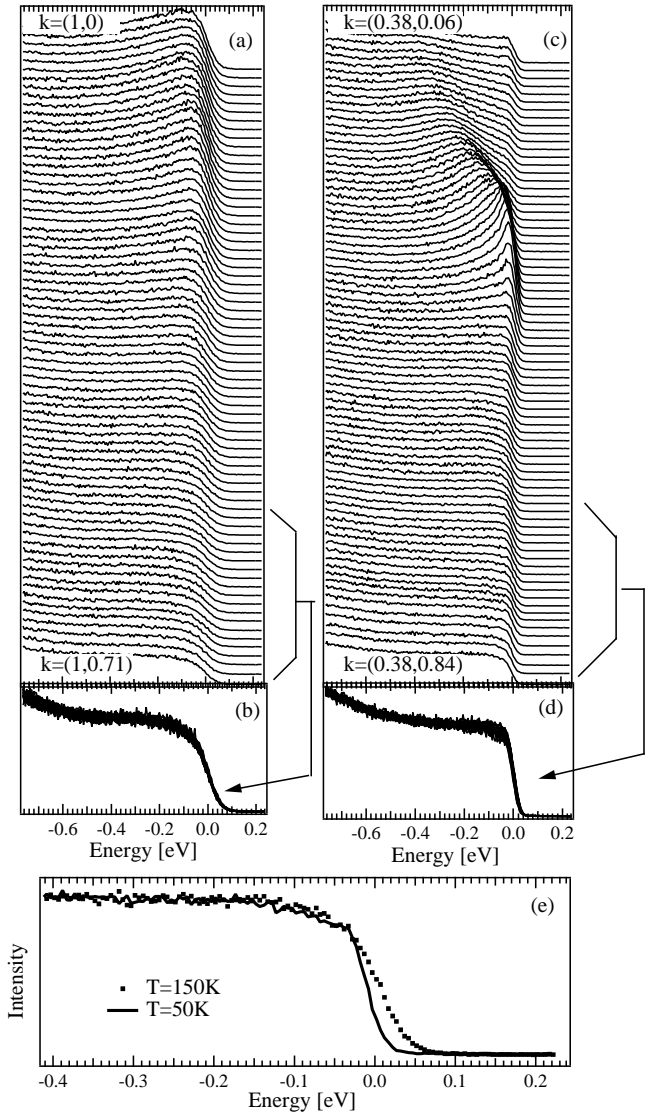}
\caption{\label{fig1} EDCs from an optimally doped film:
(a) along $(1,0)-(1,1)$ direction (in $\pi/a$ units)at T=200K,
(b) overlay of the 20 lowest EDCs from panel (a),
(c) parallel cut through the nodal Fermi point $(0.38,0.38)$ at T=50K,
(d) overlay of the 20 lowest EDCs from panel (c).
(e) background spectra at k=$(0.45,0.45)$ in normal (150K) and superconducting (50K) states, showing the shift of the leading edge towards higher binding energies in the superconducting state.  k labels in panels (a) and (c) are in $\pi/a$ units.
}
\end{figure}

We show data illustrating the background signal in Fig.~1. Panels (a)  and (c) display ARPES data going from the occupied states (top) to well beyond the Fermi momentum (k$_{F}$) along the the $(1,0)-(1,1)$ direction and a parallel cut through the nodal Fermi point $(0.38,0.38)$ respectively (momentum values are in $\pi/a$ units). The spectra far beyond k$_{F}$ show very little momentum dependence, yet they retain quite a significant intensity compared to the spectra inside the Fermi surface. This can easily be seen by examining panels (b) and (d), where we show the bottom 20 energy distribution curves (EDCs) from each cut.
The background lineshape strongly resembles that of the Fermi function, including its temperature dependence.
The data in panel (b) were taken at 200K, while data in panel (d) 
were measured at 50K. One can easily see that the width of the background leading edge increases with temperature.
Interestingly, the midpoint of this leading edge in the superconducting state is shifted from the chemical potential towards higher binding energy by approximately the same amount as the midpoint shift of the antinode (maximum superconducting gap) spectra at the Fermi momentum. This occurs even along the diagonal direction where the gap in the spectral function at k$_{F}$ (the node) is zero (Fig.~1e).

\begin{figure}
\includegraphics[width=3.5in]{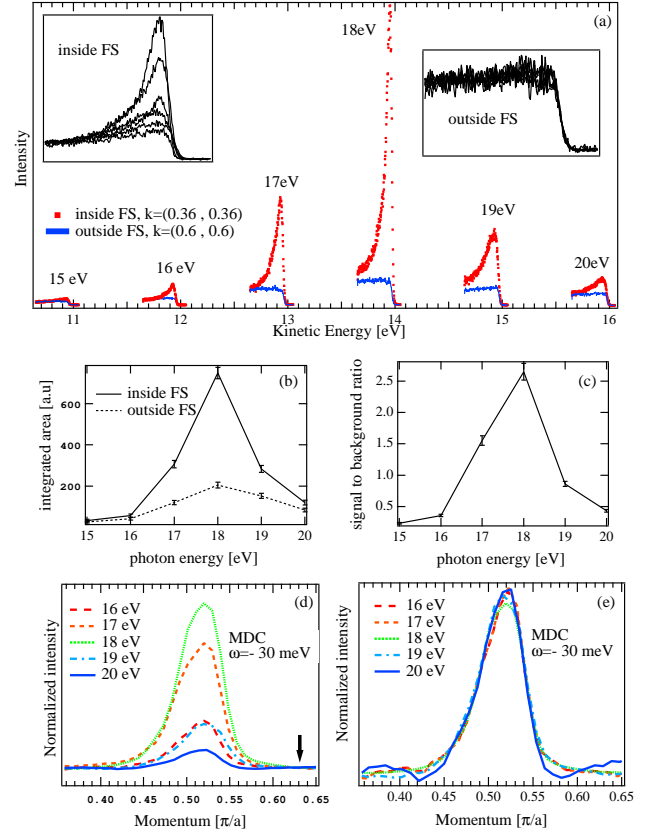}
\caption{\label{fig2} Photon energy dependence of the data for an optimally doped single crystal: 
(a) EDCs inside and outside the Fermi surface at various photon energies. Insets show the spectra inside and outside of the Fermi surface normalized at high binding energy,
(b) photon energy dependence of the energy integrated intensity from (a) for spectra inside and outside the Fermi surface,
(c) photon energy dependence of the ratio of signal to background.  k labels in panel (a) are in $\pi/a$ units.
(d) Photon energy dependence of the momentum distribution curves along the nodal direction at $\omega=-30$ meV normalized at the k point indicated by the arrow.
(e) Same data as in panel d, after background subtraction and normalization to the area under the curve.
}
\end{figure}

To test whether the observed background is an intrinsic part of the spectral function, we utilize the fact that spectra representing a single spectral function scale in exactly the same way at different k-points as a function of photon energy. We have therefore performed measurements at various photon energies and compare spectra measured inside the Fermi surface with the ones well outside, which represent the background. If the background is an intrinsic part of the spectral function, then the intensities of both should scale the same way with photon energy for two sufficiently close values of momenta. This conclusion is easily drawn from inspecting the intrinsic part of the ARPES intensity:
 \begin{equation}
I=M(h\nu, k) A(k,\omega)f(\omega)
\end{equation}  
where $M(h\nu,k)$ is the dipole matrix element, $A(k,\omega)$ the spectral function, and $f(\omega)$ the Fermi function. This form is valid if there is only one energy band present along the $(0,0) - (1,1)$ direction, as expected for the cuprates. In this case, the matrix element is expected to vary slowly as a function of momentum and the lineshape of the EDCs at all k points scale the same way with photon energy, since the spectral function itself does not depend on photon energy.
If on the other hand the background is due to an extrinsic effect, spectra inside the Fermi surface do not scale with photon energy, since in this case, the ARPES intensity is given by:
\begin{equation}
I=M_{1}(h\nu, k) A(k,\omega)f(\omega) + M_{2}(h\nu, k) B(k,\omega)f(\omega)
\end{equation}
where $B(k,\omega)$ is the background function and the matrix elements $M_{1}(h\nu, k)$  
and  $M_2(h\nu, k) $ will in general depend differently on the photon energy.  

In Fig. 2a, we plot the spectra obtained at various photon energies both inside and outside the Fermi surface along the $(0,0)-(1,1)$ direction of optimally doped Bi2212. 
From these data, we conclude that the weight of an EDC is much larger inside k$_{F}$ than outside at a photon energy of 18eV, whereas they become comparable at 15 or 20 eV. 
In order to obtain a quantitative comparison, in Fig. 2b we plot directly the energy integrated intensity for both curves at each photon energy. Here one can see that the integrated intensity both inside and outside k$_{F}$ peaks at about 18 eV, indicating that the two are related, however they are not proportional to each other. 
The occurrence of the maximum in the intensity at 18 eV is known to be due to matrix elements and is in good agreement with theoretical calculations \cite{BANSIL}. To examine the relation between the two intensities 
in more detail, in Fig. 2c we plot the ratio of the signal (defined as the difference between the intensity inside and outside k$_{F}$) divided by the intensity outside k$_{F}$. Here one can clearly see that at 18 eV the ``intrinsic" signal is 2.5 times stronger than the background, while at other photon energies it constitutes only about half of the background intensity. This pronounced photon energy dependence of the ratio gives strong evidence that the background is not a part of the spectral function that gives rise to the intrinsic signal inside of the Fermi surface.
We further note that the EDCs inside the Fermi surface do not scale as a function of the photon energy, contrary to the signal outside of the Fermi surface as seen in the  insets of panel (a). This can only be explained if the ARPES intensity is a sum of two independent components as in Eq.~2.

Similar conclusions can be reached by analyzing momentum distribution curves (MDCs). In panel (d) we plot MDC data at a binding energy of 30 meV obtained at the same photon energies as in panel (a).  These data were normalized outside the Fermi surface, at the momentum indicated by the arrow, and it is easy to see that the data do not scale close to the MDC peak position. When the background is subtracted and data normalized to the area under the curves (panel e), then the MDC peak lineshape becomes independent of the photon energy, suggesting that this part of the data is indeed due to a single component spectral function.

The next question we should try to answer regards the origin of the background signal. At this stage we do not have a definite answer. However, we believe that scattering of photoelectrons is the most likely cause. These compounds have very short escape depths of 3-5 \AA \cite{MIKEESCAPE}. The CuO plane, from which electrons with energies close to the chemical potential originate, is located some 12 \AA\ \  below the sample surface. One would therefore expect a large number of photoelectrons to undergo scattering, leading to a loss of momentum information. In contrast, in the case of simple metals, the photoelectrons originate from the top most layer, therefore even if the escape depth is the same, fewer electrons will undergo scattering, leading to a much smaller background signal. This hypothesis is supported by two properties of the background: the existence of the shift of the background leading edge towards higher binding energies from the chemical potential in the superconducting state (Fig.~1e), and the fact that the maximum intensity of the background occurs at the same photon energy (18 eV) as the maximum intensity of the intrinsic signal (Fig.~2b). These observations are consistent with the scattering scenario, since the background will result from a momentum averaging of all the photoelectrons, and all electrons near the Fermi energy have roughly the same characteristics.
Therefore if a maximum in the intensity occurs at some photon energy in a specific part of the Brilloun zone due to the matrix elements, a much smaller maximum in the background intensity will be observed, as the latter involves some type of zone
averaging.
Moreover, the Fermi function like behavior of the background signal reveals that the scattering that gives rise to it is likely elastic in origin.  The fact that the background is not due to secondaries was discussed in our earlier paper \cite{MIKEESCAPE}.
\begin{figure}
\includegraphics[width=3.5in]{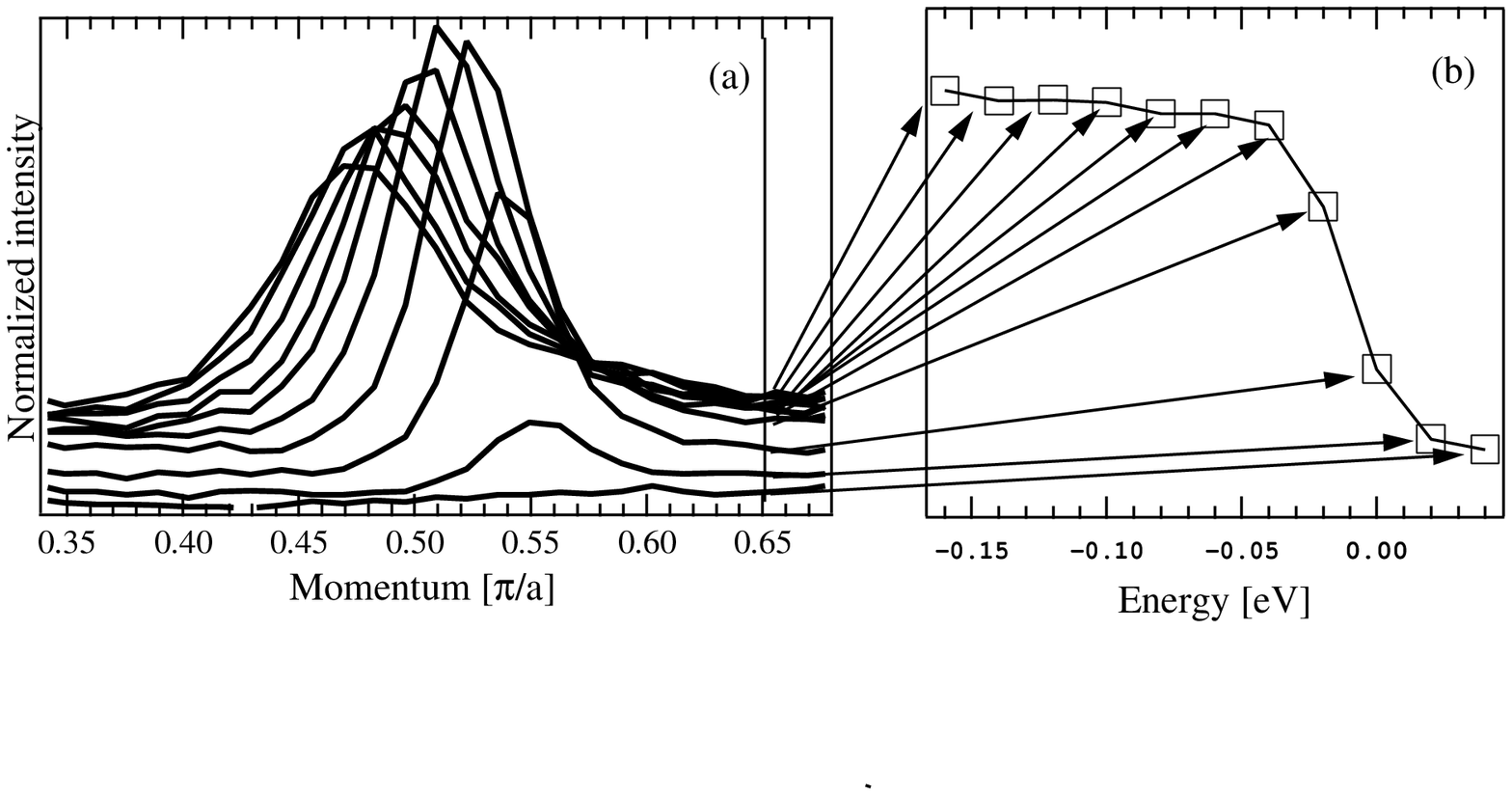}
\caption{\label{fig3} Illustration of the background subtraction procedure for an optimally doped sample at T=140K along the nodal direction:
(a) MDC data are fit by a lorentzian plus a linear background of the form ($a+bk$).
(b) intensity of the linear component is plotted as a function of energy, revealing the shape of the background at momentum vector 0.65 $\pi/a$.
}
\end{figure}

In the last part of the paper, we will concentrate on a method for subtracting the background from the data. Previously proposed methods, each with its own advantages and disadvantages, were all based on analysing  EDCs. Some involve subtracting a curve located outside and far from k$_{F}$ from all EDCs, while others treat the background as a fitting parameter \cite{HONGQP,FENGQP}. Here we present a method that takes advantage of analysing the data in terms of MDCs  \cite{JOHNSONVALLA}. In this method, the MDC lineshape is fit by a lorentzian plus a linear component of the form $a+bk$. The lorentzian is of course part of the spectral function and can be used for further analysis. Here we are interested in the linear component which constitutes the background. The method is illustrated in Fig.~3. The MDC lineshape at various binding energies is fit by a lorentzian plus a line. Then, for a given k point, the magnitude of the linear component is plotted as function of binding energy. One can easily see that this background has a shape similar to the one in Fig. 1, and also that the intensity of the background is changing slightly as a function of the momentum - it increases with increasing momentum (i.e., $b$ is non-zero). This behavior is consistent with the dipole matrix elements, which act to suppress intensity for low values of momenta.  However its precise origin is not clear and will require more theoretical work on understanding the photoelectron scattering processes. This effect also illustrates that previous methods of background subtraction were not quite accurate because of this ``hidden'' momentum dependence.

In summary, we have shown that the background present in the ARPES spectra of high temperature cuprate superconductors is extrinsic and most likely due to scattering of the photoelectrons. We have also proposed a new method for background subtraction that is more accurate than previously used methods.

This work was supported by the NSF DMR 9974401, the U.S.~DOE, Office of Science, under contract W-31-109-ENG-38, the CREST of JST, and the Ministry of Education, Science and Culture of Japan. The Synchrotron Radiation Center is supported by NSF DMR 9212658. JM is supported by the Swiss National Science Foundation, and MR in part by the Indian DST through the Swarnajayanti scheme. AK is supported by the Royal Society of Great Britain.


\begin{thebibliography}{99}

\bibitem{RMP}
For a recent review, see: A. Damascelli, Z. Hussain, and Z.-X. Shen, Rev. Mod. Phys. {\bf 75}, 473 (2003).

\bibitem{BACK}
L. Z. Liu, R. O. Anderson, and J. W. Allen, J. Phys. Chem. Solids {\bf 52}, 1473 (1991);
Z.-X. Shen and G. A. Sawatzky, Phys. Stat. Sol. (b) {\bf 215}, 523 (1999).

\bibitem{BANSIL}
S. Sahrakorpi, M. Lindroos and A. Bansil, Phys. Rev. B. {\bf 68}, 054522-1 (2003).

\bibitem{MIKEESCAPE}
M. R. Norman, M. Randeria, H. Ding, and J. C. Campuzano, Phys. Rev. B. {\bf 59}, 11191 (1999).

\bibitem{HONGQP}
H. Ding, J.R. Engelbrecht, Z. Wang, J.C. Campuzano, S.-C. Wang, H.-B. Yang, 
R. Rogan, T. Takahashi, K. Kadowaki, and D. G. Hinks,
Phys. Rev. Lett. {\bf 87}, 227001 (2001).

\bibitem{FENGQP}
D.L. Feng, D.H. Lu, K.M. Shen, C. Kim, H. Eisaki, A. Damascelli,
R. Yoshizaki, J.-I. Shimoyama, K. Kishio, G.D. Gu, S. Oh, A. Andrus,
J. O'Donell, J.N. Eckstein, Z.-X. Shen,
Science {\bf 289}, 277 (2000).

\bibitem{JOHNSONVALLA}
T. Valla, A.V. Fedorov, P.D. Johnson, B.O. Wells,
S.L. Hulbert, Q. Li, G.D. Gu, and N. Koshizuka, 
Science {\bf 285}, 2110 (1999).

\end{thebibliography}
\end{document}